\documentclass[a4paper,12pt]{article}
\usepackage{comment}
\usepackage{cite}
\usepackage{amsmath}
\usepackage{amssymb}
\usepackage{amsfonts}
\usepackage[T1]{fontenc}
\usepackage[utf8]{inputenc}
\usepackage{graphicx}
\usepackage{fancyhdr}
\usepackage{float}
\usepackage{xcolor}
\usepackage{authblk}
\usepackage{mathrsfs}
\usepackage{empheq}
\usepackage[hyphens]{url}
\usepackage{hyperref} 
\usepackage[]{breakurl}

\usepackage{geometry}
\newcommand{\threej}[6]{\left(\begin{array}{ccc}#1 & #2 & #3 \\ #4 & #5 & #6 \end{array}\right)}

\newcommand{\cle}[6]{C_{#1 #2, #3 #4}^{#5 #6}} 

\begin{document}

\title{Clebsch-Gordan coefficients, hypergeometric functions and the binomial distribution}

\author[$\dagger$]{Jean-Christophe {\sc Pain}$^{1,2,}$\footnote{jean-christophe.pain@cea.fr}\\
\small
$^1$CEA, DAM, DIF, F-91297 Arpajon, France\\
$^2$Universit\'e Paris-Saclay, CEA, Laboratoire Mati\`ere en Conditions Extr\^emes,\\ 
F-91680 Bruy\`eres-le-Ch\^atel, France
}

\maketitle

\begin{abstract}
A particular case of degenerate Clebsch-Gordan coefficient can be expressed with three binomial coefficients. Such a formula, which may be obtained using the standard ladder operator procedure, can also be derived from the Racah-Shimpuku formula or from expressions of Clebsch-Gordan coefficients in terms of $_3F_2$ hypergeometric functions. The O'Hara interesting interpretation of this Clebsch-Gordan coefficient by binomial random variables can also be related to hypergeometric functions ($_2F_1$), in the case where one of the parameters tends to infinity. This emphasizes the links between Clebsch-Gordan coefficients, hypergeometric functions and, what has been less exploited until now, the notion of probability within the framework of the quantum theory of angular momentum. 
\end{abstract}

\section{Introduction}

The Clebsch-Gordan coefficient $\cle{a}{\alpha}{b}{\beta}{c}{\gamma}$ is defined as \cite{Varshalovich1988}:
\begin{equation}
\cle{a}{\alpha}{b}{\beta}{c}{\gamma}=(-1)^{a-b+\gamma}\sqrt{2c+1}\threej{a}{b}{c}{\alpha}{\beta}{-\gamma},
\end{equation}
where $\threej{a}{b}{c}{\alpha}{\beta}{-\gamma}$ is a $3jm$ symbol. By manipulating ladder operators:
\begin{equation}
\hat{L}^{\pm}|\ell,m\rangle=\left[(\ell\mp m)(\ell\pm m+1)\right]^{1/2}|\ell,m\pm 1\rangle,    
\end{equation}
where $|\ell,m\rangle$ is an eigenvector of $\hat{L}^2$ and $\hat{L}_z$, O'Hara obtained the following expression \cite{Ohara2001}:
\begin{equation}\label{new}
\left[\cle{\frac{\ell_1}{2}}{\frac{\ell_1}{2}-k_1}{\frac{\ell_2}{2}}{\frac{\ell_2}{2}-k_2}{\frac{\ell}{2}}{\frac{\ell}{2}-k}\right]^2=\frac{\displaystyle\binom{\ell_1}{k_1}\displaystyle\binom{\ell_2}{k_2}}{\displaystyle\binom{\ell}{k}}.
\end{equation}
Let us consider the Racah-Shimpuku formula \cite{Racah1942,Shimpuku1960,Shimpuku1963,Rao1988,Wei1999,Ozay2023}:

\begin{equation}
\cle{a}{\alpha}{b}{\beta}{c}{\gamma}=\delta_{\gamma,\alpha+\beta}\left[\frac{\binom{2a}{a+b-c}\binom{2b}{a+b-c}}{\binom{a+b+c+1}{a+b-c}\binom{2a}{a-\alpha}\binom{2b}{b-\beta}\binom{2c}{c-\gamma}}\right]^{1/2}\sum_z(-1)^z\binom{a+b-c}{z}\binom{a-b+c}{a-\alpha-z}\binom{b+c-a}{b+\beta-z},
\end{equation}
with $a=\ell_1/2$, $b=\ell_2/2$ ($\ell=\ell_1+\ell_2$), $c=\ell/2$, $\alpha=\ell_1/2-k_1$, $\beta=\ell_2/2-k_2$, $\gamma=k-\ell/2$ (with $k=k_1+k_2$). Only the $z=0$ term contributes in the sum. Thus, one gets immediately Eq. (\ref{new}), the result obtained by O'Hara.

In Ref. \cite{Varshalovich1988}, several formulas are given pp. 240 and 241 which relate the Clebsch-Gordan coefficient to $_3F_2$ hypergeometric series. Let us take the first of them (Eq. (21), p. 240); the  result (\ref{new}) can also be obtained from the latter expression:
\begin{align}
C_{a\alpha,b\beta}^{c\gamma}=&\delta_{\gamma,\alpha+\beta}\frac{\Delta(abc)}{(a+b-c)!(-b+c+\alpha)!(-a+c-\beta)!}\left[\frac{(a+\alpha)!(b-\beta)!(c+\gamma)!(c-\gamma)!(2c+1)}{(a-\alpha)!(b+\beta)!}\right]^{1/2}\nonumber\\
&\times~_3F_2\left[
\begin{array}{c}
-a-b+c,-a+\alpha,-b-\beta\\
-a+c-\beta+1,-b+c+\alpha+1
\end{array};1\right].
\end{align}
Setting $-a-b+c=0$, $-a+\alpha=-k_1$, $-b-\beta=k_2$, $-a+c-\beta+1=k_2+1$ and $-b+c+\alpha+1=\ell_1-k_1+1$, the latter hypergeometric function boils down to
\begin{equation}
~_3F_2\left[
\begin{array}{c}
0,-k_1,k_2\\
k_2+1,\ell_1-k_1+1
\end{array};1\right]=1
\end{equation}
since $(0)_p=\delta_{0,p}$ and $(q)_0=1$ $\forall (p,q)$, where $(a)_k=a(a+1)(a+2)\cdots(a+k-1)=\Gamma(a+k)/\Gamma(a)$ is the Pochhammer symbol and $\delta_{p,q}$ the Kronecker symbol. The coefficient $\Delta(abc)$  is defined as
\begin{equation}
\Delta(abc)=\left[\frac{(a+b-c)!(a-b+c)!(-a+b+c)!}{(a+b+c+1)!}\right]^{1/2}.
\end{equation}
In the present case, since $a+b-c=0$, $a-b+c=\ell_1$, $-a+b+c=\ell_2$ and $a+b+c+1=\ell+1$, one gets
\begin{equation}
\Delta(abc)=\left[\frac{\ell_1!\ell_2!}{(\ell+1)!}\right]^{1/2} 
\end{equation}
and concerning the remaining term, since $a+b-c=0$, $-b+c+\alpha=\ell_1-k_1$ and $-a+c-\beta=k_2$, we find
\begin{equation}
\frac{(a+\alpha)!(b-\beta)!(c+\gamma)!(c-\gamma)!(2c+1)}{(a-\alpha)!(b+\beta)!}=\frac{(\ell_1-k_1)!k_2!k!(\ell-k)!(\ell+1)}{k_1!(\ell_2-k_2)!}
\end{equation}
and
\begin{equation}
(a+b-c)!(-b+c+\alpha)!(-a+c-\beta)!=(\ell_1-k_1)!k_2!
\end{equation}
yielding
\begin{align}
\cle{\frac{\ell_1}{2}}{\frac{\ell_1}{2}-k_1}{\frac{\ell_2}{2}}{\frac{\ell_2}{2}-k_2}{\frac{\ell}{2}}{\frac{\ell}{2}-k}=&\left[\frac{\ell_1!\ell_2!}{(\ell+1)!}\right]^{1/2}\frac{1}{(\ell_1-k_1)!k_2!}\left[\frac{(\ell_1-k_1)!k_2!k!(\ell-k)!(\ell+1)}{k_1!(\ell_2-k_2)!}\right]^{1/2}\\
=&\left[\frac{\displaystyle\binom{\ell_1}{k_1}\binom{\ell_2}{k_2}}{\displaystyle\binom{\ell}{k}}\right]^{1/2},
\end{align}
which is the expected result.

\section{The binomial distribution as a limit of a hypergeometric function}

The binomial $(n,p)$ distribution is the limit of the hypergeometric $(n_1,n_2,n_3)$ distribution with $p=n_1/n_3$ as $n_3\rightarrow\infty$. Let the random variable $x$ have the hypergeometric $(n_1,n_2,n_3)$ distribution. The probability mass function of $x$ is

\begin{align}
    \mathscr{P}(x=n_1)=&\frac{\displaystyle\binom{n_1}{x}\binom{n_3-n_1}{n_2-x}}{\displaystyle\binom{n_3}{n_2}}\\
    =&\frac{n_1!}{x!(n_1-x)!}\frac{(n_3-n_1)!}{(n_2-x)!\left[(n_3-n_1)-(n_2-x)\right]!}\frac{n_2!(n_3-n_2)!}{n_3!}\\
    =&\frac{n_2!}{x!(n_2-x)!}\frac{n_1!(n_3-n_1)!(n_3-n_2)!}{(n_1-x)!(n_3-n_1-n_2+x)!n_3!}\\
    =&\binom{n_2}{x}\left[n_1(n_1-1)\cdots (n_1-x+1)\right]\\
     &\times\frac{\left[(n_3-n_1)(n_3-n_1-1)\cdots (n_3-n_1-n_2+x+1)\right]}{n_3(n_3-1)\cdots (n_3-n_2+1)!},
\end{align}
for $x=0, 1, 2, \cdots, n_2$. The hypergeometric distribution can also be defined through its probability generating function $G(t)$:
\begin{equation}
    G(t)=\frac{\displaystyle\binom{n_3-n_1}{n_2}}{\displaystyle\binom{n_3}{n_2}}\,~_2F_1\left[\begin{array}{c}
    -n_1,-n_2\\
    n_3-n_1-n_2+1\\
    \end{array};t\right]
\end{equation}
or through its moment generating function $M(t)$:
\begin{equation}
    M(t)=\frac{\displaystyle\binom{n_3-n_1}{n_2}}{\displaystyle\binom{n_3}{n_2}}\,~_2F_1\left[\begin{array}{c}
    -n_1,-n_2\\
    n_3-n_1-n_2+1\\
    \end{array};e^t\right].
\end{equation}
The expectation value of the hypergeometric distribution is
\begin{equation}
    \mathbb{E}[X]=n_1n_2/n_3
\end{equation}
and its variance 
\begin{equation}
    \mathrm{Var}[X]=\frac{n_1n_2(n_3-n_1)(n_3-n_2)}{n_3^2\left(n_3-1\right)}.
\end{equation}
Since $n_1=pn_3$ and $n_3\rightarrow\infty$, one has also $n_1\rightarrow\infty$. We expect that $n_3-n_1\leq n_3-n_2$ and $n_2$ can be ignored as $n_1, n_3$ tend to infinity. Setting $\xi=1/p=n_3/n_1$, one gets, for the probability $\mathscr{P}$ that variable $x=n_1$:
\begin{align}
    \mathscr{P}(x=n_1)=&\binom{n_2}{x}\left[n_1(n_1-1)\cdots (n_1-x+1)\right]\\
    &\times\frac{(\xi-1)n_1\left[(\xi-1)n_1-1\right]\cdots\left[(\xi-1)n_1-n_2+x+1\right]}{\xi n_1(\xi n_1-1)\cdots (\xi n_1-n_2+1)}\\
    =&\binom{n_2}{x}\left(\frac{1}{\xi}\right)^x\left(\frac{\xi-1}{\xi}\right)^{n_2-x}\\
    =&\binom{n_2}{x}p^x\left(1-p\right)^{n_2-x},
\end{align}
with $x=0, 1, 2,\cdots, n_2$, which is the probability mass function for the binomial $B(n_2,x,p)$ distribution, where
\begin{equation}
    B(q,r,p)=\binom{q}{r}p^r(1-p)^{q-r}.
\end{equation}
Let us consider, following O'Hara \cite{Ohara2001}, $K_1$ and $K_2$ independent binomial random variables with distributions $B(\ell_1,k_1,p)$ and $B(\ell_2,k_2,p)$ respectively and $m_i=\ell_i/2-k_i$. The sum of two independent binomial random variables with common parameter $p$ is itself a binomial random variable of parameter $p$ \cite{Bickel1977}. Indeed, if $K_1$ and $K_2$ are binomial random variables with the moment generating function $\left[pe^t+(1-p)\right]^{\ell_1}$ and $\left[pe^t+(1-p)\right]^{\ell_2}$ respectively, then the moment generating function of $K=K_1+K_2$ is a binomial random variable with binomial distribution $B(\ell=\ell_1+\ell_2,k,p)$. One has
\begin{equation}
    \mathscr{P}\left(m=\frac{\ell}{2}-k\right)=\mathscr{P}\left(\frac{\ell}{2}-K=\frac{\ell}{2}-k\right)=\mathscr{P}(K=k)=\binom{\ell}{k}p^k(1-p)^{\ell-k}
\end{equation}
and the conditional probability
\begin{align}
    &\mathscr{P}\left(m_1=\frac{\ell_1}{2}-k_1, m_2=\frac{\ell_2}{2}-k_2\,\Big|\,m=\frac{\ell}{2}-k\right)=\frac{\displaystyle \mathscr{P}\left(m_1=\frac{\ell_1}{2}-k_1, m_2=\frac{\ell_2}{2}\right)}{\displaystyle \mathscr{P}\left(m=\frac{\ell}{2}-k\right)}\\
    &=\frac{\displaystyle\binom{\ell_1}{k_1}p^{k_1}(1-p)^{\ell_1-k_1}\binom{\ell_2}{k_2}p^{k_2}(1-p)^{\ell_2-k_2}}{\displaystyle\binom{\ell}{k}p^{k}(1-p)^{\ell-k}}\\
    &=\frac{\displaystyle\binom{\ell_1}{k_1}\displaystyle\binom{\ell_2}{k_2}}{\displaystyle\binom{\ell}{k}}=\left[\cle{\frac{\ell_1}{2}}{\frac{\ell_1}{2}-k_1}{\frac{\ell_2}{2}}{\frac{\ell_2}{2}-k_2}{\frac{\ell}{2}}{\frac{\ell}{2}-k}\right]^2.
\end{align}

\section{Conclusion}

In this note, we showed how the special Clebsch-Gordan coefficient discussed by O'Hara can be obtained from the Shimpuku-Racah expansion or from expressions in terms of the $_3F_2$ hypergeometric function of unit argument. Since O'Hara also exhibited an interesting relation with the binomial probability distribution, we make the connection between the Gauss hypergeometric function $_2F_1$ distribution by considering the binomial distribution as one of its limits. We believe that the links between Clebsch-Gordan coefficients, hypergeometric functions, and the probabilistic approach are likely to enlighten some aspects of the quantum theory of angular momentum \cite{Bailey1935,Andrews1999,Cantarini2022}.

\end{document}